\documentclass[10pt,conference]{IEEEtran}

\usepackage{cite}
\usepackage{amsmath,amssymb,amsfonts}
\usepackage{graphicx}
\usepackage{textcomp}
\usepackage{float}
\usepackage{multirow}
\usepackage{multicol}
\usepackage{amsmath}

\newcommand{\inlineeqnum}{\refstepcounter{equation}~~\mbox{(\theequation)}}

\graphicspath{ {res/} }
\begin{document}

\title{Robust and Scalable Techniques for TWR and TDoA based localization using Ultra Wide Band Radios}

\author{
    \IEEEauthorblockN{Rakshit Ramesh\IEEEauthorrefmark{1}, Aaron John-Sabu\IEEEauthorrefmark{2}, Harshitha S\IEEEauthorrefmark{3}, Siddarth Ramesh\IEEEauthorrefmark{1},\\ Vishwas Navada B\IEEEauthorrefmark{1}, Mukunth Arunachalam\IEEEauthorrefmark{1}, Bharadwaj Amrutur\IEEEauthorrefmark{1}}
    \IEEEauthorblockA{\IEEEauthorrefmark{1}Robert Bosch Center for Cyberphysical Systems, Indian Institute of Science, Bangalore \\ Email: rakshitr@iisc.ac.in}
    \IEEEauthorblockA{\IEEEauthorrefmark{2}Department of Electrical Engineering, Indian Institute of Technology Bombay, Mumbai \\ Email: aaronjohnsabu1999@gmail.com}
    \IEEEauthorblockA{\IEEEauthorrefmark{3}Department of Electrical Engineering, Indian Institute of Technology Madras, Chennai \\ Email:s.harshitha016@gmail.com }
}

\maketitle

\begin{abstract}
Current trends in autonomous vehicles and their applications indicates an increasing need in positioning at low battery and compute cost.
Lidars provide accurate localization at the cost of high compute and power consumption which could be detrimental for drones.
Modern requirements for autonomous drones such as No-Permit-No-Takeoff (NPNT)  and applications restricting drones to a corridor require the infrastructure to constantly 
determine the location of the drone.
Ultra Wide Band Radios (UWB) fulfill such requirements and offer high precision localization and fast position update rates at
a fraction of the cost and battery consumption as compared to lidars and also have greater network availability than GPS in a dense forested campus or an indoor setting. 
We present in this paper a novel protocol and technique to localize a drone for such applications using a Time Difference of Arrival (TDoA) approach.
This further increases the position update rates without sacrificing on accuracy and compare it to traditional methods.
\end{abstract}

\section{Introduction}
Localization involves finding the position of the target in a coordinate system, mostly Euclidean. The coordinate system maybe spanned as the Global Positioning coordinate system
or a relative one. 
Conventional localization using GPS follows a principle of Time difference of Arrival of signals coming in from 4 or more satellites.
Given that the satellites also broadcast their position (latitude, longitude and elevation),
the receiver can compute its position which is a function of the time differences of a packet transmitted by the satellites \cite{gps}.
This procedure assumes that the satellites have their clocks synchronized and the receiver is also able to synchronize its own clock w.r.t the clock of the satellites.
Satellites have a very precise atomic clock and drift very little. This luxury is not available for use in a similar UWB-based localization scheme which we aim to be low cost.
In addition, GPS also has the limitation of satellite signal reception indoors and in environments with a canopy, where Ultra Wide Band Radios can be used instead.

The radio is capable of precisely transmitting messages and recording timestamps on received messages.
This allows us to perform a simple Time of Flight  or Time Difference/Phase Difference of Arrival based measurements.
We can localize a radio module using either the Time of Flight or \textit{Two Way Ranging} (TWR) based approach or using the \textit{Time Difference of Arrival} (TDoA) approach.

The manufacturer of the module supports a TWR scheme where the time of flight of a packet determines the range. This method generally requires a larger number of messages
to determine the range \cite{dwmum}.
An alternate method is the TDoA scheme which relies on the disparity in arrival time of a message from the target to receiving stations.
This method relies heavily on all radios of the system to be synchronized.
Prior literature on the subject in \cite{robusttdoa} and \cite{tdoa1} describes a technique of performing TDoA without quantifying real world performance.
\cite{selfloco} describes a station initiated technique of TDoA for drone odometry purposes.
Our work primarily describes the TWR technique and its limitations and we present our approach to the TDoA problem with the target initiating the localization.
We believe our approach is more scalable in terms of the ease of synchronizing clocks, having a large deployment area and being able to localize more number of targets,
usually in large industrial warehouses and other GPS denied environments primarily for infrastructure monitoring of drones.

UWB radios spread information over a large band, greater than 500MHz. The IEEE 802.15.4a UWB describes the specification for the UWB impulse radio physical layer.
The radio in use, the DWM1000 \cite{dwmum} implements the specification and operates in the band between 3.5GHz to 6.5GHz.
The modulation scheme employs Burst Position Modulation (BPM) to locate start of data portions which are modulated using Binary Phase Shift Keying (BPSK).
The on-board clock is a 38.4 MHz reference crystal upscaled to 63.8976 GHz (approximated to 64 GHz; i.e. 15.65 ps time period) using internal phase-locked loops (PLLs).

Henceforth, we shall refer to the stationary beacon as \textit{Anchor} and the beacon on a moving drone as a \textit{Tag}.

\section{Two Way Ranging (TWR)}
\begin{figure}[h]
\centering
\includegraphics[scale=0.5]{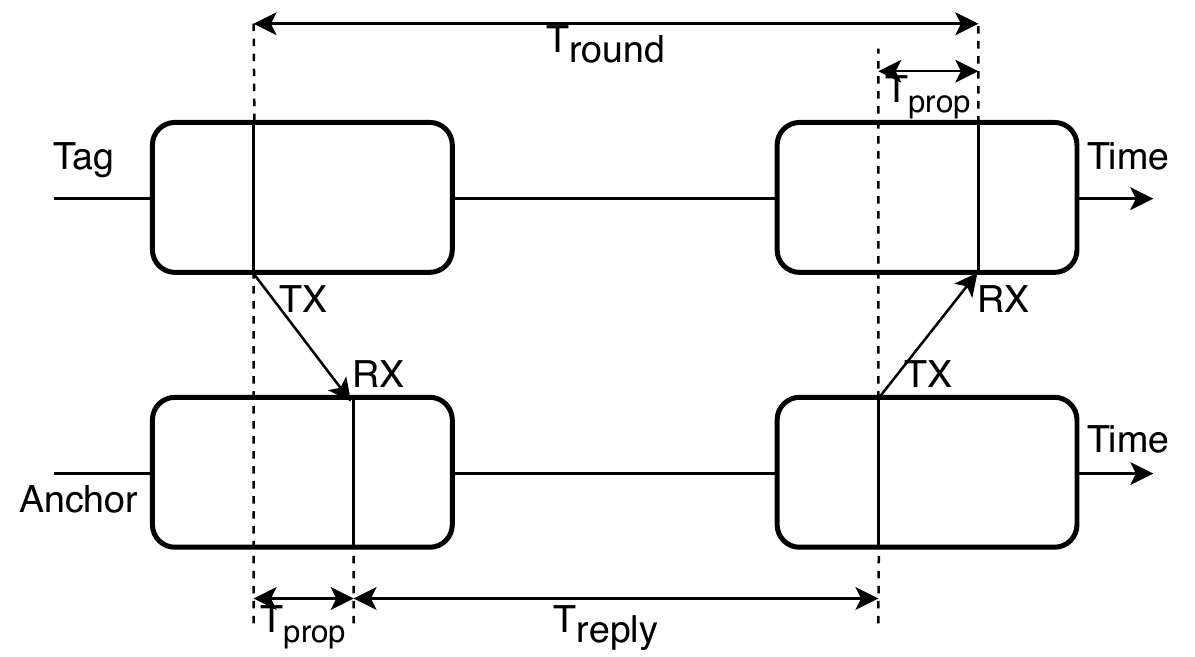}
\caption{Two Way Ranging timing sequence diagram}
\label{fig:twr_seq}
\end{figure}

Two Way Ranging involves a \textit{tag} transmitting a POLL to an \textit{anchor} which responds back with a POLL-ACK acknowledgement message. 
Both the tag and anchors record the timestamps at which a packet was transmitted/received.
Fig.\ref{fig:twr_seq} shows a single round messaging sequence diagram. The time of flight $\hat{T}_{prop} = (T_{round} - T_{reply})/2$ where $T_{round}$ is the time difference between the
tag transmitting a POLL message and receiving the POLL-ACK and $T_{reply}$ is the time difference between the
anchor receiving the POLL message and transmitting the POLL-ACK message. The range  $R$ can be now determined from $T_{prop}$ assuming the message lengths in both cases are same.

This scheme relies on inherent time delays between the receiver and the transmitter which is not known before hand and which depends heavily on the channel of propagation.
Hence, it is more useful to employ a \textit{Symmetric Double-Sided Two-Way Ranging scheme} involving two rounds of messaging and is described in \cite{dwmum}.
We find from our experiments with ranging at a fixed distance $d_A$ that the ranging so obtained falls along a normal distribution centered around $d_A + \epsilon_A$ for some anchor \textit{A},
where $\epsilon_A$ is a bias for this anchor-tag pair and can be subtracted out from subsequent measurements.

\begin{figure}[h]
\centering
\includegraphics[scale=0.4]{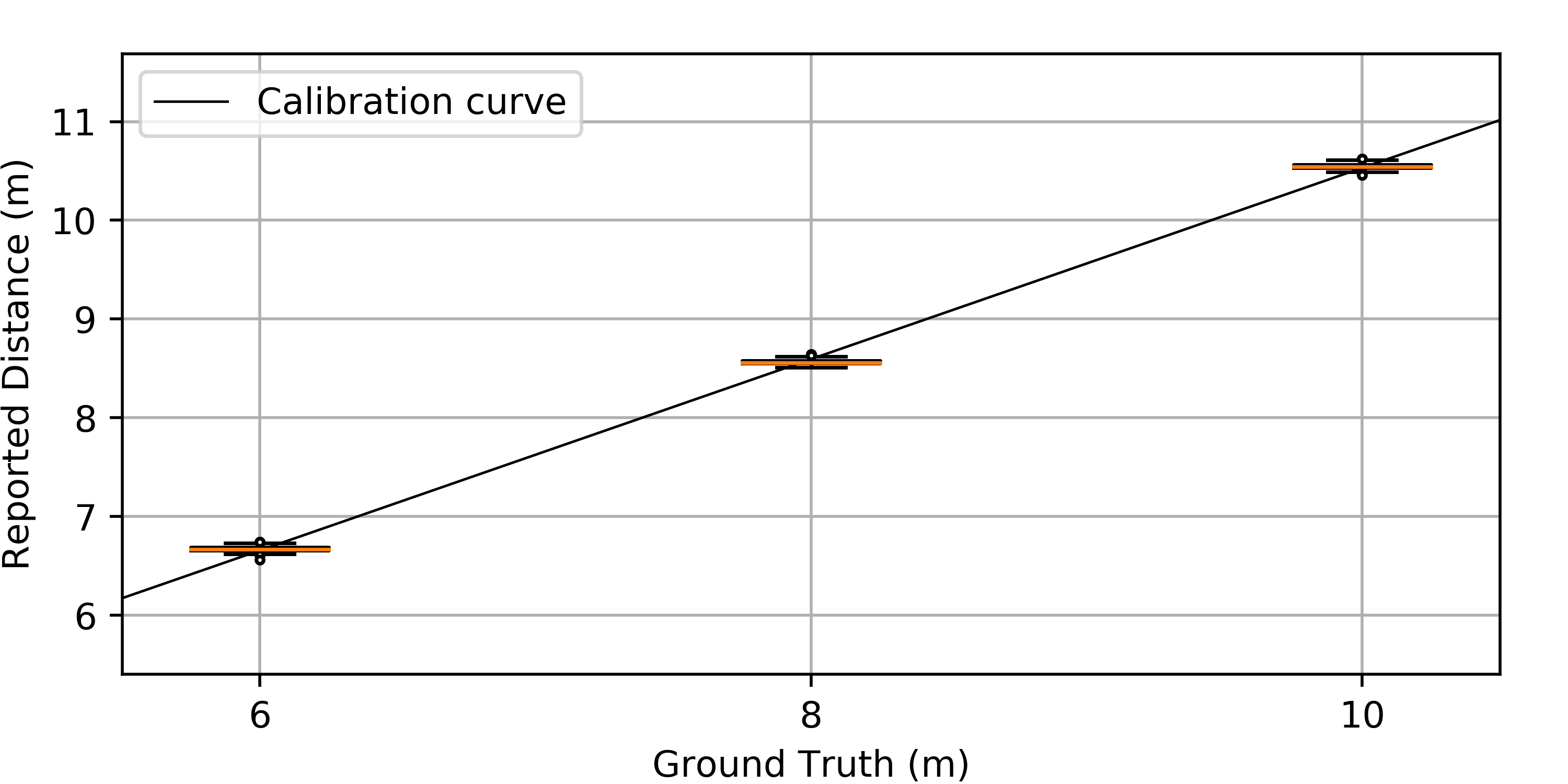}
\caption{A calibration plot for a pair of tag and anchors with box plot for measured points. This shows an offset of $ \epsilon_a = 0.847m$ and slope $ m = 0.96$ }
\label{fig:1d}
\end{figure}

A calibration step will be required to find $\epsilon_A$ by measuring range for different distance.
We perform a simple linear regression as shown in Fig.\ref{fig:1d} to find the offset and subsequently subtract it from all measurements.

\subsection{Localization based on Two Way Ranging}
A simple 2-D Localization of the \textit{tag} can be obtained by trilateration.
In 2-D Euclidean space, a simple closed form solution for the Cartesian position $p_t = (x, y)$  coordinates can be obtained by solving the equation,
$ r_k^2 = (x_k - x)^2 + (y_k - y)^2 $ $k \in$ anchors $(A,B,C)$ and $r_k$ is the measured radial distance between that respective anchor $k$ and the tag \inlineeqnum\label{eqn:clf}.
This is a problem of finding the point of intersection of three circles (for each tag-anchor pair) and has been discussed in detail in \cite{closed}.
However, the closed form solution is not easily extensible when more redundancies in the form of extra anchors are available.
Including additional anchors also affords a more precise solution by decreasing the uncertainty in localization because of noisy measurements \cite{noisyloco}.
Hence, we treat trilateration as a convex optimization problem as explained in \cite{locopt} and \cite{twropt}. The position of the drone can be obtained by solving (\ref{eqn:twrcost}).
\begin{equation}
    p_T = \underset{p_T}{argmin} \sum_{k} \ r^2_k - \| p_k - p_T \|_{l2}^2, k \in (A,B,C)
\label{eqn:twrcost}
\end{equation}

We have experimented a few convex optimization techniques, the Powell Method, BFGS Method and the Non-Linear Least Squares method with Trust Region Reflective Method
and compared them in Table.\ref{tbl:twr_conv}.
We find the localization accuracy in terms of percentage of the number of measured positions
which are within a radius of the actual position and the time taken per position update in Table.\ref{tbl:twr_conv}.
We also note the execution time for finding the solution (Core-i5 laptop).

\begin{table}[H]
\caption{\label{tbl:twr_conv} Comparing different methods for solving 2-D TWR based localization}
\begin{center}
\begin{tabular}{|c|c|ccc|}
\hline
Anchors              & Method             & Time          & $\leq$ 10cm         & $\leq$ 15cm \\
\hline
  3                  & Closed Form        & $2.1e-6$      & 90                  & 99.14       \\
                     & Powell Method      & $3.4e-3$      & 70                  & 98.4        \\
                     & BFGS               & $7.2e-4$      & 70                  & 98.4        \\
                     & Least Squares (TRF)& $5e-3$        & 70                  & 98.11       \\
\hline
  2                  & Closed Form        & -             & -                   & -           \\
                     & Powell Method      & $4e-3$        & 54.13               & 94.81       \\
                     & BFGS               & $6e-4$        & 20                  & 36.4        \\
                     & Least Squares (TRF)& $2e-3$        & 30                  & 51.1        \\
\hline
\end{tabular}
\end{center}
\end{table}

There are several advantages of using an iterative solution to localization. For e.g, we can incorporate multiple anchor measurements into getting more accurate results.
As shown in Table.\ref{tbl:twr_conv}, we can also perform 2-D localization with 2 anchors. This is possible by incorporating bound constraints and history (last localized point)
in finding the solution. Also, because measurements from these sensors are noisy, it is possible that a closed form solution will lead to unstable solutions.
We see that for two anchors only, the Powell Method \cite{powell} performs best but takes the most time because of greedy search and offers little convergence guarantees.
BFGS \cite{bfgs} method takes the least time.

\subsection{Drawbacks of Localization based on Two Way Ranging}
\label{sec:twrdraw}
In the symmetric double-sided two-way ranging scheme, the number of anchors required to fully localize the tag and disambiguate quadrants is $n_{anc} = n_{dim} + 1$
where $n_{dim}$ is 2 or 3 for 2-D and 3-D space respectively.
If the tag initiates ranging with an anchor, the computed range will only be available with the anchor which will then have to re-transmit the range to the tag.
If the anchor initiates ranging, a \textit{time division mulitplexing scheme} amongst the anchors of the system will have to be employed to avoid packet collision.
Ranging with each tag-anchor involves a total of $n_{msgs} = 3$ messages (POLL, POLL-ACK, RANGE) for anchor initiated ranging and $n_{msgs} + 1 = 4$ for tag initiated ranging.
In our case, with an absence of an anchor synchronization scheme, we chose tag initiated ranging.
Therefore, 2-D localization will require $n_{msgs} * n_{anc} = 12$ total messages between all the anchors and the tag for successful localization.
An Anchor node can only reply at around $500\mu s$ after receiving a message.
This severely impacts the total localized position refresh rate of the tag to around $ 6ms $ and therefore the position update rate becomes around $12Hz$
The drone moves at a maximum velocity of $2m/s$ and therefore at $12Hz$ position refresh, the drone would have already moved by $16cms$
before its position updates. This can be critical for certain tasks such as precision landing.
Therefore, a scheme which allows a faster position update is required.

\section{Time Difference of Arrival (TDoA) based Localization}
The TWR mode of ranging relies on the time of flight of a message from the transmitter to the receiver.
In contrast, the Time Difference of Arrival (TDoA) mode of ranging relies on the disparity in the time of arrival of a message.
As mentioned in Sec.\ref{sec:twrdraw}, TWR supports low position update rates owing to the interplay of signalling message sequences. TDoA relies on fewer messages
for localization and scales well in real world scenarios where a large number of drones need to be simultaneously localized.
However, TDoA requires that the clocks of all the anchors are synchronized.

\subsection{Clock drift}
\begin{figure}[h]
\centering
\includegraphics[scale=0.5]{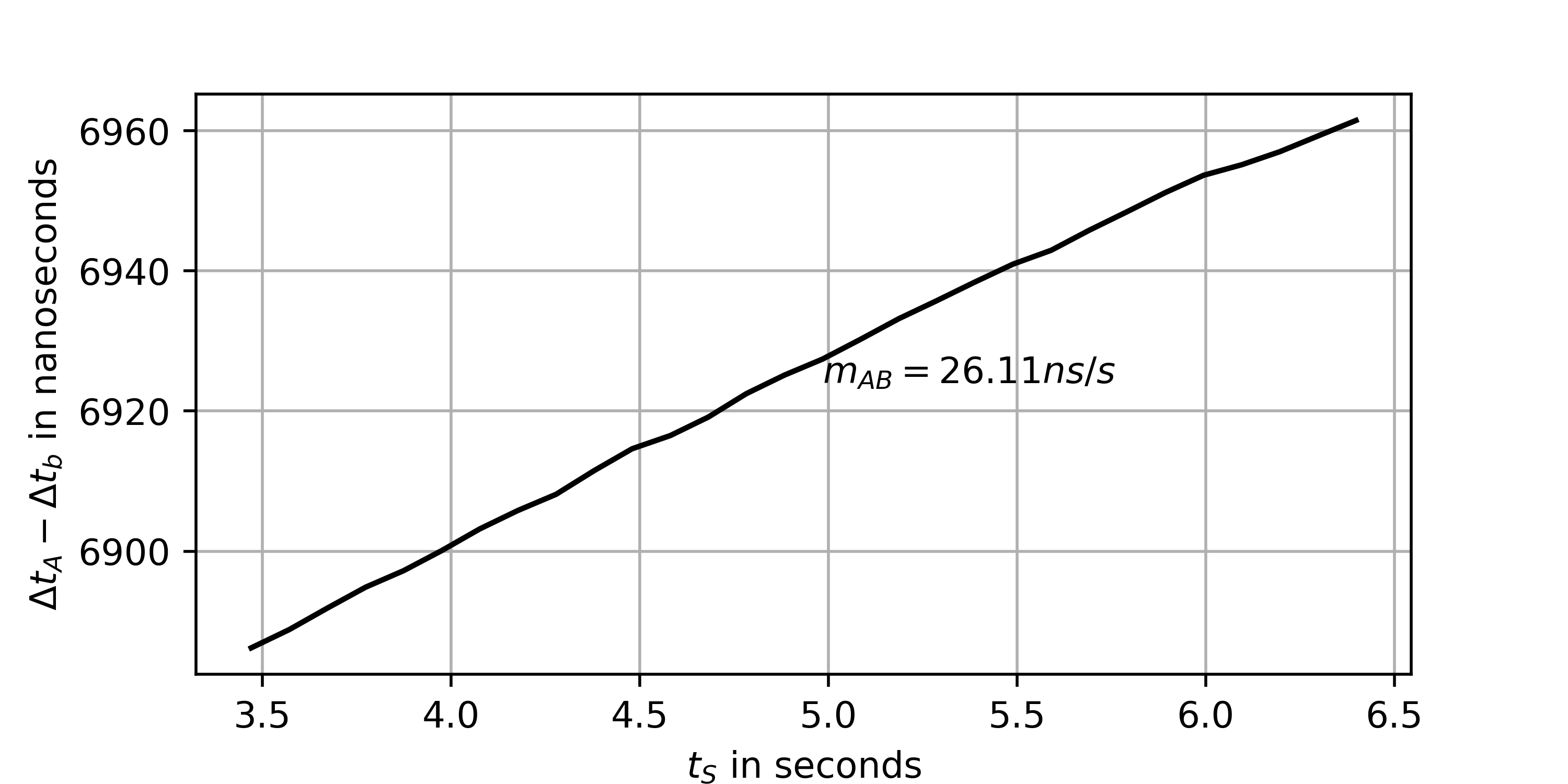}
\caption{Time drift between two nodes at a sampling rate of 1Hz}
\label{fig:clock_diff}
\end{figure}

Because of the lack of two way messaging involved, TDoA based localization heavily relies on the clocks of the transmitter and receiver being in synchronization.
Owing to the manufacturing inaccuracies in designing the radios and their clocks, the clock frequency differs over time.
Consider two nodes, A placed at $p_A = (0, 0)$ and B placed at $p_B = (0, 4)$.
A synchronization node S is placed at $p_s = (2, 0)$ and periodically broadcasts a SYNC message at a rate $f_{sync}$.
Node S measures $t_{S,i}^{(tx)}$ for every $i^{th}$ epoch. Nodes A and B measure the time at which they receive a SYNC message from node S $t_{k, i}^{(rx)}, k \in (AB)$.
We relate the time difference between nodes A or B receiving and S transmitting two consecutive (i and i-1) SYNC messages as
$\Delta{t}_{AB} = m_{AB} \times \Delta{t}_{s}$ between the $i^{th}$ and $(i-1)^{th}$ epochs. 
We record the timestamps $t_{S, i}^{(tx)}$ at which the $i^{th}$ SYNC message was broadcast and
$t_{k,i}^{(rx)}$ the time at which the broadcast was received by $k \in (AB)$.
Fig.\ref{fig:clock_diff} shows $\Delta{t}_{A} - \Delta{t}_{B}$ vs $t_{S, i}$ for $f_{sync} = 1Hz$.
Because Nodes A and B drift at different rates, we would expect a straight line with slope $m_{AB}$.
We observer that $m_{AB} = 26 ns/s$, i.e for every $1s$ sampling interval of Node S, the observed intervals of Nodes A and B differ by $26 ns$.
In $1ns$, light travels approximately $0.3m$, therefore, in a $1s$ interval, the errors in the time difference of arrival of a packet from Node S to Nodes A and B would be $8m$.
For a different synchronization period  $f_{sync} = 20Hz$, we obtain a drift rate $m_{AB} = 13.6 ns/s$ corresponding to a drift of 4m every $1s$.
From this experiment we find that it is impossible to synchronize nodes by simply reseting their clocks on every reception of a SYNC message and that the UWBs only provides good accuracy 
as a short interval timer.

\subsection{Time difference of Arrival Calculation}
\label{sec:tdoa_calc}
\begin{figure}[h]
\centering
\includegraphics[width=\linewidth]{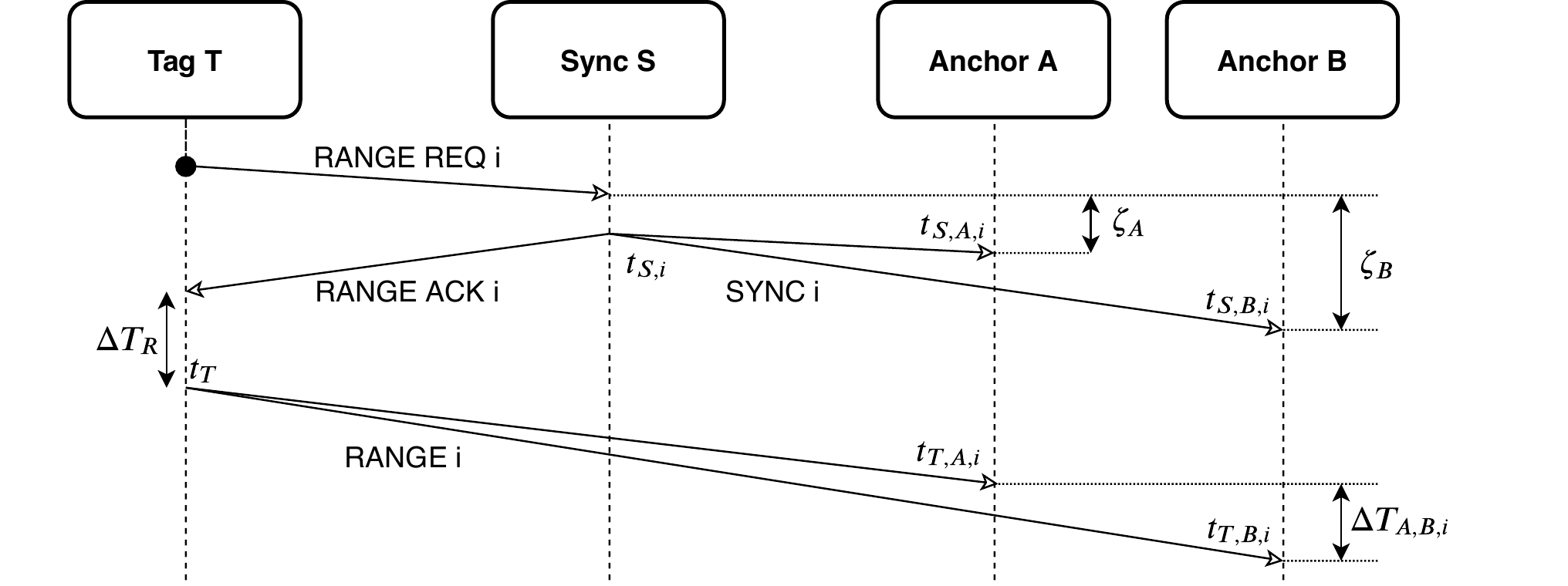}
\caption{Time Difference of Arrival Message Sequence Diagram}
\label{fig:tdoa}
\end{figure}

\begin{figure}[h]
\centering
\includegraphics[width=\linewidth]{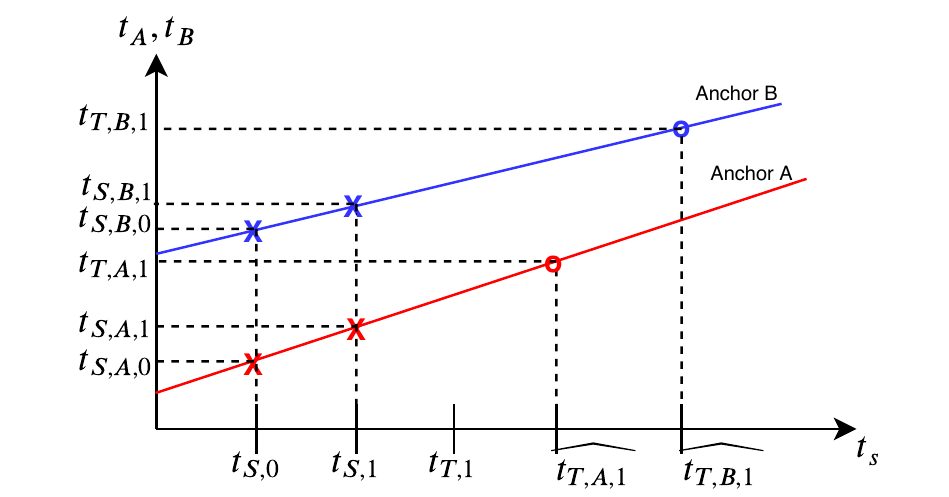}
\caption{Timing diagram illustrating the sequence of messages and their respective Received/Transmitted times}
\label{fig:tdoa_time}
\end{figure}

There are two submodes in TDoA, forward mode - where a Tag broadcasts a marker packet and the time difference of arrival amongst the anchors is measured.
In the reverse mode, the anchors synchronously broadcast marker packets and the tag computes the time difference of arrival of these packets.
The reverse mode of localization is more scalable when there are more Tags to be localized, however,
it is harder to implement owing to the need for synchronizing all the anchors to a base clock.
The reverse mode also suffers from low localization refresh rates because of the need to time multiplex anchors broadcasting their marker packets.
Considering these, we have devised a protocol which simplifies the anchor synchronization process and performs localization in forward TDoA mode.
Fig.\ref{fig:tdoa} shows a message sequence diagram of the protocol.
The setup is as follows - There are three Anchors A, B and C and one Synchronization Node S at positions $(x_k,y_k), k \in (A,B,C,S)$.
Nodes A, B and C power on at different times and have different initial values and their clocks count at different rates.
A signal transmitted by Node S reaches Nodes $(A,B,C)$ in time $(\zeta_{A},\zeta_{B},\zeta_{C})$ respectively. Since we know the exact distance 
of the Sync Node S from the Anchor, this can be calculated as $\zeta_k = d_k/c \ k \in (A,B,C)$.
A tag T at an unknown position $(x_T,y_T)$ is in the vicinity of this setup and is in communicable distance with all nodes.
For the sake of explanation, let's consider only Nodes A and B and the sequence of events at instant $i$.
Localization is initiated by the Tag by first requesting the Sync node (S) with a "RANGE REQ" message. Node S then broadcasts a "SYNC" message.
The anchors receive these messages at $t_{S,A}$ and $t_{S,B}$ (in their respective clocks).
The same "SYNC" message is received by Tag T as an acknowledgement to proceed with the next message.
After a fixed amount of time $\Delta{T_R}$ the Reply Delay Time, the Tag T transmits a "RANGE" message. The anchors receive this at time $t_{T,A}$ and $t_{T,B}$.
This process repeats for every epoch $i \in (0,1,2...)$. A graphical representation of the timing is as shown in Fig.\ref{fig:tdoa_time}.
The x-axis represents the Sync Nodes clock and the Y axis represents the clocks of Anchor A and B (the y-axis is for representation purposes only).

Because the clocks of Nodes A and B aren't synchronized, finding the time difference of arrival of "RANGE" message between Anchors A and B will be invalid.
From Fig.\ref{fig:tdoa_time} we have two "SYNC" transmitted timestamps from the Sync Node $t_{S,i-1}$ and $t_{S,i}$
and two "SYNC" received timestamps $t_{S,k,i-1}$ and $t_{S,k,i}$
$\ k \in (A,B)$ for $i^{th}$ epoch. We can now estimate the clocks counting rate for that epoch w.r.t the Sync Nodes clock as -
$m_{k,i} = (t_{S,k,i} - t_{S,k,i-1})/(t_{S,i}-t_{S,i-1}) \ k \in (A,B)$\inlineeqnum\label{eqn:skewcalc}. The Anchors distance offset from Sync node $\zeta_{k}$ gets cancelled out from subtraction.
The Tag now transmits a "RANGE" message to obtain its position. This message is received by the Anchors at different times depending on their distance from the Tag
and their clocks starting value and drift rate. In their local clocks, these are $t_{T,A,i}$ and $t_{T,B,i}$ during the $i^{th}$ epoch.
We now find the "adjusted" time on the Sync Nodes clock when the reception of "RANGE" message was recorded on each Anchors node as shown in (\ref{eqn:adjusted_time}).

\begin{equation}
\begin{split}
\widehat{t_{T,k,i}} = & (t_{T,k,i} - t_{S,k,i} + \zeta_k)/m_{k,i} + t_{S,i} \\
                      & k \in (A,B,C) \ for \ the \ i^{th} \ epoch 
\end{split}
\label{eqn:adjusted_time}
\end{equation}

Note that the distance offset $\zeta_k$ for messages received from Sync node needs to be subtracted from $t_{S,k,i}$.
We can now find the adjusted time difference of arrival of a "RANGE" packet between two Anchors as - 
\begin{equation}
\begin{split}
    \Delta{T_{k,l,i}} = & \widehat{t_{T,k,i}} - \widehat{t_{T,l,i}} \\
                        & (k,l) \in (A,B,C)\ \land k \neq l for \ the \ i^{th} \ epoch 
\end{split}
\label{eqn:timediff}
\end{equation}

\subsection{Clock interpolation using Kalman Filters}
We see from our algorithm that every Tag's "RANGE" message needs to be preceded by two "SYNC" messages to synchronize the anchors and estimate the clock drift before ranging.
This is needed because from \ref{fig:clock_diff} we see that the UWBs clock is effective only as a short interval timer and the anchor clocks drift apart from each other very quickly. 
This requirement can be averted by using a Linear Kalman filter for estimating clock skews and correctly estimating the Tags "RANGE" messages time of arrival on the anchor.
A UWB Anchor nodes clock can be modeled as $t_{S,k,i} = t_{k,i-1} + m_{k,i}\Delta{t_{S,i}}$
where $\Delta{t_{S,i}}$ is the time interval between two transmitted "SYNC" messages, $t_{S,k,i}$ is the Anchor Nodes clock reading when it received the $i^{th}$ "SYNC" message
and $m_{k,i}$ is the Anchor nodes clock skew at instant $i$.
Consider $x_{k,i} = [t_{S,k,i}, m_{k,i}]^T$ to be the state vector for an Anchor $k$.
For every reception of "SYNC" message, we predict the current state as -
$ x_{k,i}' = \begin{bmatrix} 1 & 0\\ 0 & 1 \end{bmatrix}x_{k,i-1} + \begin{bmatrix} 0 & 1\\ 0 & 0 \end{bmatrix}[\Delta{t_{S,i}}, 0]^T$
We then update the estimate by computing the residual as -
$ y_{k,i} = [1,0] [t_{S,k,i}, m_{k,i}]^T $ where $t_{S,k,i}$ is the internal measurement of the Anchors received timestamp and $ m_{k,i}$ can be found from (\ref{eqn:skewcalc}).
The state $x_{k,i}$ can then be obtained using standard Kalman Filter equations.
We need to carefully choose the Process Covariance Matrix $P$ and the Measurement Noise Matrix $R$.
We choose $\sigma^2_{t} = 0.4ns^2$ and $\sigma^2_{m} = 0.01$ through analyzing timing records.
An initial value of $P= \begin{bmatrix} 1 & 0 \\ 0 & 0.001 \end{bmatrix}$ works well in making the filter give equal weightage to predictions and measurements.
With this filtered estimate of $t_{S,k,i} \ and \ m_{k,i}$ we can find the filtered and adjusted time difference of arrival of a message as described in (\ref{eqn:adjusted_time}).

\subsection{Localization from TDoA}
We can now localize the Drone (Tag T) and obtain its cartesian coordinates assuming that the Anchors have been synchronized.
In \cite{robusttdoa} the authors combine the clock skew problem and the localization problem as a single state estimation problem but the authors don't quantify 
performance with real world data.
We take a more simplified approach where we ensure clock drifts are mitigated instead 
of incorporating position estimations into our state equations and solve for the position separately.
We find that \cite{selfloco} have a similar approach as us except that they operate in the reverse mode described in Sec.\ref{sec:tdoa_calc}.
Our technique operates in the forward mode and finds its application in tracking the drone to ensure permitted flight.
We have Anchor nodes $k$ placed at $p_k = (x_k,y_k)$. Assume the Tag Node on the drone is currently at position $p_T = (x_t,y_t)$. 
At instant $i$ the Tag T transmitted a "RANGE" message at time $T_{T,i}$.
For instant i we have the \textit{time difference of arrivals} $\Delta{t_{k,l,i}} $ for $(k,l) \in (A,B,C) \land k \neq l $ from (\ref{eqn:timediff}).
For each anchor pair the following holds approximately true - 
$ c \times \Delta{t_{k,l,i}} =  \|p_T - p_k\| + \|p_T - p_l\| $ where $ (k,l) \in (A,B,C) \land k \neq l $ and $c$ is the speed of light.
We can solve for the drones position $p_T$ iteratively as shown in (\ref{eqn:cost}).
\begin{equation}
    \begin{split}
        p_{T,i} = \underset{p_{T,i}}{argmin}\ \sum_{(k,l)} c \Delta{t_{k,l,i}} & - \|p_{T,i} - p_k\|_{l2}  + \|p_{T,i} - p_l\|_{l2} \\
                                                        & (k,l) \in (A,B,C) \ \land \ k \ \neq l
    \end{split}
\label{eqn:cost}
\end{equation}
Essentially, this process is solving for the point of intersection of three hyperbolas.
We use non-linear least squares method using the Trust Region Reflective Algorithm \cite{trust} to solve for (\ref{eqn:cost}).

\subsection{Localization performance}
We establish a system of UWB nodes with three Anchor nodes (A,B,C), one Synchronization node S and one Tag node T in a room measuring $8m \times 8m$.
Anchor nodes A,B and C are placed at $ (5.2,4.3), (0,0) \ and \ (0, 4.3)$, Sync node S is at $ (2,0) $. The Tag node T is placed at various positions inside the room.
We evaluate the performance of the algorithm with and without a Kalman filter and for different synchronization intervals and find the percentage of points (out of 1000) 
which lie withing $20cm$ and $10cm$ as shown in Table.\ref{tbl:tdoa_comp}.

\begin{table}[H]
\caption{\label{tbl:tdoa_comp} Performance of algorithm for various synchronization intervals}
\begin{center}
\begin{tabular}{|c|c|ccc|}
\hline
Tag Position         & Method                 & Sync            & $\leq$ 20cm         & $\leq$ 10cm \\
\hline
(0,2)                & No filter              & $100ms$          & 91.8                & 58.1        \\
                     & No filter              & $300ms$          & 84.8                & 47.6        \\
                     & No filter              & $500ms$          & 60                  & 30          \\
                     & With filter            & $100ms$          & 91.8                & 58.2        \\
                     & With filter            & $300ms$          & 90.5                & 57          \\
                     & With filter            & $500ms$          & 65                  & 30          \\
\hline
\end{tabular}
\end{center}
\end{table}

Table.\ref{tbl:tdoa_comp} shows the localization algorithms performance for different "Sync" synchronization intervals. This essentially describes the 
amount of time between two synchronization messages sent by the Sync Node after which a Tag node sends a "RANGE" request.
We see from the table that the algorithm succeeds in localizing to within 20cm of the target point and offers $92\%$ accuracy when the last synchronization was $100ms$ prior.
As the time since the last synchronization increases to about $500ms$ we see a gradual deterioration in performance.

As noted in previous sections, TDoA requires fewer ($3$ messages in TDoA vs $12$ in TWR) messaging signals and therefore offers a better position update rate.
The UWB module allows the Tag node to send a
reply message within $500\mu s$ of receiving a "SYNC" message from a sync node. Assuming at worst, each "RANGE" message requires two "SYNC" messages, we will have 3 messages per ranging.
Therefore we can theoretically achieve around $600Hz$ position update rates. However, the accuracy is observed to be much better in the TWR case. Experiments we have performed
that are not mentioned in this paper show us that this can be bettered by incorporating
drone local sensor measurements such as an Inertial Measurement Unit \cite{imu}, camera or an optical flow sensor. 
Owing to limitations in taking this information out of the radio and on to an aggregator through serial UART interface (RaspberryPi3), we achieve around $50Hz$
position update rates which is a significant improvement over TWR method ($10Hz$). 
There is also an inherent scalability advantage in TDoA because of lesser channel utilization owing to fewer messages being passed around.
In \cite{selfloco} mentions spatially varying measurement biases in TDoA based localization. In our experiments, we have found the same to be true in some cases owing to multipath fading. 
However, in a large region of observation within the spanned area of the anchors, our algorithm 
yields similar performances and deteriorates only when the Tag is very close ($\leq 50cm$) to an anchor.
Using fast Internet Transport protocols like MQTT as shown in \cite{mqlat},
we can reliably transfer the localized point from the infrastructure back to the drone in  case it needs it.

\section{Conclusions}
We have compared Two Way Ranging based localization to our novel Time Difference of Arrival based localization with clock synchronization.
Our technique provides fast position update rates and managed to do so with a simple clock synchronization scheme.
Using our technique, we have shown that almost $91\%$ of the localized positions can be made to fall within a 20cm radius of the actual position and 
we can do so by synchronizing only once every $100ms$. This would enable high speed positioning that Drones require and we are able to localize a node at a $50Hz$ rate. 
Going forward, we are experimenting with using Gaussian Processes to learn how the clock departs from its expected behaviour and therefore have a better estimate of the drift before
estimating the time of arrival of a message. We are also experimenting with quantifying the spatial variation of biases which occur due to multipath fading and factor it into our localization technique.

\bibliography{./references}
\bibliographystyle{ieeetr}

\end{document}